# Atomic bonding and electrical characteristics of two-dimensional graphene/boron nitride van der Waals heterostructures with manufacturing defects via binding energy and bond-charge model

Jiannan Wang[#], Liangjing Ge[#], Anlin Deng, Hongrong Qiu, Hanze Li, Yunhu Zhu,

Maolin Bo*

Key Laboratory of Extraordinary Bond Engineering and Advanced Materials Technology (EBEAM) of Chongqing, Yangtze Normal University, Chongqing 408100, China

[#]These authors contributed equally to this work.

*Corresponding Author: E-mail: bmlwd@yznu.edu.cn (Maolin Bo)

**Abstract**

We used the binding energy and bond-charge model to study the atomic bonding and electrical properties of the two-dimensional graphene/BN van der Waals heterostructure. We manipulated its atomic bonding and electrical properties by manufacturing defects. We discovered that this process yielded a band structure with a flat band, i.e., a horizontal band structure without dispersion at the Fermi level. Thus, our research is significant because it is the first report on this flat band of defect graphene/BN van der Waals heterostructures.

**Keywords**: graphene/BN heterostructures, flat band, structural defects

.

# 1. Introduction

In 2004, Geim et al. succeeded in experimentally preparing graphene for the first time via mechanical exfoliation[1]. They consequently proved that two-dimensional (2D) materials can stably exist. Graphene has excellent properties such as high strength[2], high thermal conductivity, and high electron mobility[3]. it is also not highly sensitive to temperature changes, even at high temperatures. The stable existence of graphene is attributable to 1) its reaction to thermal conduction kinetic energy, which induces out-of-plane deformation, and 2) the fact that all of the carbon (C) atoms of graphene are not strictly uniformly distributed in a 2D plane. These factors reduce the energy of graphene[4, 5]. Similarly, the discovery of graphene has allowed researchers to discover more 2D materials[6], such as boron nitride (BN), transition metal disulfides, tungsten disulfide, and MXene materials[7, 8]. Its dimensionality reduction process also affords graphene many novel characteristics[9], such as the quantum spin Hall effect[10], quantum superconductivity[11], and the exciton effect[12], which are different from the optoelectronic properties of bulk materials. It can be applied in optoelectronic devices, chemical environments, and fields related to energy and batteries[10, 13-15].

Various 2D materials can selectively grow and stack on substrates to form a heterojunction[16]. Two-dimensional heterojunction materials are layered materials that rely on the occurrence of van der Waals interactions between layers[17]. Different prototype materials and stacking methods can be applied to ensure that the resulting structure has the desired shape and properties. Although different materials and methods have been used, all of the processes entailed the application of 2D materials were also the same[18]. Additionally, different 2D materials can be grown and vertically stacked on different substrates. This is possible because there is a van der Waals force between the 2D materials that joins them and allows for the construction of a new type of 2D van der Waals heterojunction[19]. Typically, the process of stacking 2D materials can yield surprising physical and chemical properties[20]. Studies have shown that 2D van der Waals heterojunctions not only possess the properties of the original parent material, but also that the effects of stacking tend to cause the 2D material to demonstrate characteristics that are unique from those of the original single-layer material, which broadens the spectrum of possibilities[21-23]. Similarly, although it is an insulator, boron nitride(BN) has a band gap of 6.64 eV. The lattice matching of the heterojunction structure is less than 1.14%. Thus, BN can be considered to be the best substrate for graphene.

The defects of interface atoms are known to significantly alter the physical properties of crystalline materials. Their effects primarily affect the morphology, strength, band gap, and electrochemical properties, such as the dielectric constant and magnetic moment[24, 25]. According to the principle of thermodynamic equilibrium, there is no perfect crystal, and all materials have defects[26]. In most processes of material synthesis and growth, the defect density, defect type, and defect morphology can be controlled to a certain extent by controlling the temperature, pressure, and other experimental conditions[27]. The introduction of defects typically traps the free carriers of the material near the defect, thereby triggering the localization of phonons, excitons, and charges[28]. Thus, the existence of line defects significantly impacts the mechanical properties of 2D materials[29]. Han et al. found that, in the armchair direction, the tensile strength at the graphene grain boundary increases with increasing orientation angle, whereas the tensile strength at the graphene grain boundary in the zigzag direction increases non-monotonously[30]. Bohayra et al. found that, as the concentration of defects increases, the elastic modulus, tensile strength, and failure strain of graphene tend to decrease[31]. Furthermore, the ability of graphene to conduct heat is more sensitive to the occurrence of defects than the mechanical properties.

In this study, we constructed a graphene/BN van der Waals heterojunction with BN as the substrate. We were consequently able to obtain different structures by manufacturing surface defects. The performances of the electrical and chemical bonds were regulated, and we applied density functional theory (DFT) and binding energy and bond-charge (BBC) model to study the defective graphene/BN heterojunctions. Additionally, we analyzed the band gap, atomic bonding, deformation charge densities, density of states (DOS), and various other parameters of all of the manufactured structures. We hope that this method of optimizing the structural performance via the manufacture of defect heterojunctions can be applied in future studies.

## 2. Principles

### 2.1 DFT calculations

In this study, we applied a projection-based plane-wave DFT method, which entailed the implementation of Cambridge Sequential Total Energy Package (CASTEP) for atomic-scale material simulation. We also applied General Gradient Approximate (GGA)-based Perdew-Burke-Ernzerhof (PBE) [32]functionals for the calculations. Taking into consideration the interaction between the layers in the structure, we set the vacuum layer to 16 Å. To begin, we applied CASTEP to optimize the structure and adjust the position and stability of the atoms. **Table 1** lists the specific parameters of the initial structure, the interlayer spacing of which was 4.187 Å. As previously mentioned, van der Waals interactions are required for the formation of heterogeneous structures. Thus, we established our model to consider the existence of van der Waals forces. Atoms are typically joined by covalent bonds in a double-layer heterostructure, and the structural stability of the layers is dependent on the van der Waals forces. In consideration of this, we applied a dispersion-corrected method to correct the PBE functionals and thus stabilize the influence of the van der Waals forces. The cut-off energy for the plane-wave basis set was set to be 517 eV, and the k-point grid dimensions were 4×4×1. During the simulation, the energy converged to $10^{-6}$ eV, and the force applied to each atom converged to <0.01 eV/Å.

### 2.2 Binding energy and bond-charge (BBC) model

We confined the charge $e$ in one dimension and subjected it to periodic boundary conditions, i.e., those corresponding to a particle on a ring. Thus, the Hamiltonian operator of the system takes the following form:

$$H = -\frac{\hbar^2}{2m}\left(\nabla - \frac{q}{\hbar c}A(\vec{r})i\right)^2$$
$$= \frac{1}{2}\left(-i\partial_\mu - qA(\vec{r})\right)^2 \quad (e=1, \hbar=1, c=1)$$

(1)

where $\hbar$ is the Planck's constant, $q$ is the electric charge, $c$ is the speed of light, r is the electron radius, and $\psi$ is the field. We consider the transformation of the particle field $\psi \rightarrow \exp(iq\theta)\psi$. This transformation represents a symmetry of the free action of the particle if $\theta$ is a constant, but we want to consider a generic function $\theta(r)$. The $\theta(r)$ is local phase transformation. At the same time, it is known that the action of the free electromagnetic field is invariant under the following gauge transformation [33]:

$$A_\mu \rightarrow A_\mu - \partial_\mu \theta.$$

(2)

It is then possible to replace, in the action, the derivative $\partial_\mu$ with a covariant derivative of $\psi$ as

$$D_\mu \psi = (-i\partial_\mu - A_\mu)\psi,$$

(3)

So that

$$D_\mu \psi \rightarrow e^{iq\theta} D_\mu \psi$$

(4)

even when $\theta$ depends on $r$.

Considering a particle of mass $m$ with a geometrical coordinate $r$ and a potential $V_{cry}(r_{ij})$, the Hamiltonian of such a particle interacting with the electromagnetic field can be written as

$$\vec{H}^{gi} = \frac{1}{2}\left(-i\partial_\mu - qA(\vec{r})\right)^2 + V_{cry}(\vec{r})$$
$$= \frac{1}{2}\left(-i\partial_\mu - qA(\vec{r})\right)^2 + \Delta qA(\vec{r}).$$

(5)

It turns out that the expectation values $\langle \psi_v | \vec{H}^{gi} | \psi_v \rangle$ are invariant under local phase transformations,

$$\psi_v(x) \rightarrow e^{iq\theta(x)}\psi_v(x),$$

(6)

This is due to the contribution of the gauge field $A(\vec{r})$. The subscript $v$ represents the orbit of the electron.

From the energy band theory and binding energy (BE) model[34], the formula can be obtained:

$$\begin{cases} \overline{H}^{gi} = \frac{1}{2}\left(-i\partial_\mu - qA(\vec{r})\right)^2 + \Delta qA(\vec{r}) = \left[-\frac{\hbar^2 \nabla^2}{2m} + V_{atom}(\vec{r})\right] + V_{cry}(\vec{r})(1+\Delta_H) \\ E_v(0) = -\langle v,i| -\frac{\hbar^2 \nabla^2}{2m} + V_{atom}(\vec{r})|v,i\rangle \\ E_v(x) - E_v(0) = -\langle v,i|V_{cry}(\vec{r})|v,i\rangle \left[1 + \frac{\sum_j f(k)\langle v,i|V_{cry}(\vec{r})|v,j\rangle}{\langle v,i|V_{cry}(\vec{r})|v,i\rangle}\right] \\ \qquad = \alpha_v(1 + \sum_j \frac{f(k)\cdot \beta_v}{\alpha_v}) \cong \alpha_v(1+\Delta_H) \propto \langle E_x \rangle \\ \alpha_v = -\langle v,i|V_{cry}(\vec{r})|v,i\rangle \propto \langle E_b \rangle;\ \beta_v = -\langle v,i|V_{cry}(\vec{r})|v,j\rangle; \Delta_H = \sum_j \frac{f(k)\cdot \beta_v}{\alpha_v} \cong \delta\gamma \end{cases}$$

(7)

The $E_v(x)$ and $E_v(0)$ are the $v$th energy levels of the crystal atoms and an isolated atom, respectively. The $V_{cry}(\vec{r})$ is the potential energy of the crystal lattice. The $\alpha_v$ and $\beta_v$ contributes to the width of the energy band. In the localized band of core levels, $\beta_v$ is very small, so $\alpha_v$ determines the energy band of the core levels. The $\langle v,i|$ and $\langle v,j|$ represents the wave function. The periodic factor $f(k)$ is the form of $e^{ik\vec{r}}$, while $k$ is the wave vector. The $\beta$ is dependent on the overlap between orbitals centered at two neighboring atoms.

Combining the BE model with the "initial and final state effects" model[35, 36], we obtain the formula:

$$\begin{cases} V_{cry}(\vec{r})(1+\Delta_H) = \gamma V_{cry}(\vec{r}) = (Z+1)\frac{1}{4\pi\varepsilon_0}\sum_i(-\frac{e^2}{2|\vec{r}-\vec{r}'|}) \\ E_v(x) - E_v(0) \cong -\langle v,i|V_{cry}(\vec{r})(1+\Delta_H)|v,i\rangle = -(Z+1)\frac{1}{4\pi\varepsilon_0}\sum_i \langle v,i| -\frac{e^2}{2|\vec{r}-\vec{r}'|}|v,i\rangle \propto \langle E_x \rangle \\ E_v(x) - E_v(B) \cong -\delta\gamma \langle v,i|V_{cry}(\vec{r})|v,i\rangle \cong -\sum_j f(k)\langle v,i|V_{cry}(\vec{r})|v,j\rangle \end{cases}$$

(8)

Z is the atomic charge parameter. The $\gamma = Z+1$ is binding energy ratio and $\delta\gamma = \gamma - 1$ is relative binding energy ratio. $B$ represents the bulk atoms. From the above formula, we can obtain the energy-level shift formula of the charge effect:

$$\begin{cases} Z=-1(\delta\gamma=-1,\gamma=0,Nonbonding), \Delta E_v(x)=\Delta E_v(0)=0 & \text{(isolated atom, neutral atom)} \\ Z=0(\delta\gamma=0,\gamma=1,Bonding), \Delta E_v(x)=\Delta E_v(B)=-(-\sum_i \langle v,i|\frac{1}{4\pi\varepsilon_0}\frac{e^2}{2|\vec{r}-\vec{r}'|}|v,i\rangle)>0 & \text{(bulk atoms, charged positive atoms)} \\ Z=|\delta\gamma|(\delta\gamma>0,Bonding), \Delta E_v(x)=-(-\sum_i \langle v,i|\frac{(1+|\delta\gamma|)}{4\pi\varepsilon_0}\frac{e^2}{2|\vec{r}-\vec{r}'|}|v,i\rangle)>0 & \text{(charged positive atoms)} \\ Z=-|\delta\gamma|(\delta\gamma<0)\begin{cases} -1<\delta\gamma<0, Nonbonding, \Delta E_v(x)=-(-\sum_i \langle v,i|\frac{(1-|\delta\gamma|)}{4\pi\varepsilon_0}\frac{e^2}{2|\vec{r}-\vec{r}'|}|v,i\rangle)>0 \\ \delta\gamma<-1, Antibonding, \Delta E_v(x)=-(-\sum_i \langle v,i|\frac{(1-|\delta\gamma|)}{4\pi\varepsilon_0}\frac{e^2}{2|\vec{r}-\vec{r}'|}|v,i\rangle)<0 \end{cases} & \text{(charged negative atoms)} \end{cases}$$

.

(9)

The $\varepsilon_0$ is the dielectric constant of vacuum. We can calculate the chemisorption or defect-induced binding energy ratio $\gamma$ with the known reference value of $\Delta E_v'(x) = E_v(x) - E_v(B)$, $\Delta E_v(B) = E_v(B) - E_v(0)$ and $\Delta E_v(x) = E_v(x) - E_v(0)$ derived from the surface via DFT calculations and X-Ray Photoelectron Spectroscopy(XPS) analysis. Hence, we obtain

$$\delta\gamma = \frac{\Delta E_v'(x)+\Delta E_v(B)}{\Delta E_v(B)} - 1 = \gamma - 1 .$$

(10)

Thus, one can drive the relative BE ratio $\delta\gamma$. If $\delta\gamma < 0$, the BE is reduced (atom gains electron), the potential of the crystal and the bond is weakened. Conversely, if $\delta\gamma > 0$, the BE increases (atom loses electrons), the potential of the crystal and the bond becomes stronger.

The atomic bond relaxation can be represented by the bond-charge (BC) model[37], as follows:

$$\begin{cases} E_x \propto V_{cry}(\vec{r}_x) = \Delta q A(\vec{r}_x) \\ \frac{E_x}{E_b} = \left(\frac{1/d_x}{1/d_b}\right)^m \propto \frac{V_{cry}(\vec{r}_x)}{V_{cry}(\vec{r}_b)} = \gamma; \begin{cases} \gamma>1, deepening\ the\ potential\ well \\ \gamma<1, strengthing\ the\ potential\ Energy\ Barrier \end{cases} \end{cases},$$

(11)

Note that $E_x$ is the single-bond energy and $E_b$ is bond energy of the bulk atom. The $d_x$ is the bond length of the atom, and $m$ is the bond nature indicator. From the Hartree-Fock theory and the Hubbard model [38, 39], we have

$$\Delta \hat{V}_{ee} = \frac{1}{2} \int d^3r \int d^3r' \alpha_\sigma^+(\vec{r}) \alpha_\sigma(\vec{r}) \Delta V_{ee}(\vec{r}-\vec{r}') \alpha_{\sigma'}^+(\vec{r}') \alpha_{\sigma'}(\vec{r}')$$

$$= \frac{1}{2} \frac{\Delta e_r \Delta e_{r'}}{|\vec{r}-\vec{r}'|} \int d^3r \int d^3r' \alpha_\sigma^+(\vec{r}) \alpha_\sigma(\vec{r}) \alpha_{\sigma'}^+(\vec{r}') \alpha_{\sigma'}(\vec{r}')$$

$$= \frac{1}{2|\vec{r}-\vec{r}'|} \int d^3r \int d^3r' \Delta\rho(\vec{r}) \Delta\rho(\vec{r}')$$

(12)

The $\alpha_\sigma^+(\vec{r})$ is the creation operators. The $\alpha_\sigma(\vec{r})$ is the annihilation operators. For completeness, we have also endowed the electrons with a spin index, $\sigma = \uparrow/\downarrow$. The $\Delta\rho(\vec{r}) = \Delta e \rho(\vec{r}) = \Delta e \alpha_\sigma^+(\vec{r}) \alpha_\sigma(\vec{r})$ is the deformation charge density. Then, we have

$$V_{bc}(\vec{r}-\vec{r}') = \frac{1}{8\pi\varepsilon_0} \int d^3r \int d^3r' \frac{\Delta\rho(\vec{r})\Delta\rho(\vec{r}')}{|\vec{r}-\vec{r}'|}$$

($\Delta\rho_{Hole-electron} < \Delta\rho_{Antibonding-electron} < \Delta\rho_{No\ charge\ tranfer} = 0 < \Delta\rho_{Nonbonding-electron} < \Delta\rho_{Bonding-electron}$)

$$Bonding\ states \begin{cases} \Delta\rho_{Hole-electron}(\vec{r})\Delta\rho_{Bonding-electron}(\vec{r}') < 0 (Strong\ Bonding) \\ \Delta\rho_{Hole-electron}(\vec{r})\Delta\rho_{Nonbonding-electron}(\vec{r}') < 0 (Nonbonding\ or\ Weak\ Bonding) \\ \Delta\rho_{Antibonding-electron}(\vec{r})\Delta\rho_{Bonding-electron}(\vec{r}') < 0 (Nonbonding\ or\ Weak\ Bonding) \\ \Delta\rho_{Antibonding-electron}(\vec{r})\Delta\rho_{Nonbonding-electron}(\vec{r}') < 0 (Nonbonding) \\ \Delta\rho_{Nonbonding-electron}(\vec{r})\Delta\rho_{Bonding-electron}(\vec{r}') > 0 (Antibonding) \\ \Delta\rho_{Hole-electron}(\vec{r})\Delta\rho_{Antibonding-electron}(\vec{r}') > 0 (Antibonding) \\ \Delta\rho_{Hole-electron}(\vec{r})\Delta\rho_{Hole-electron}(\vec{r}') > 0 (Antibonding) \\ \Delta\rho_{Antibonding-electron}(\vec{r})\Delta\rho_{Antibonding-electron}(\vec{r}') > 0 (Antibonding) \\ \Delta\rho_{Nonbonding-electron}(\vec{r})\Delta\rho_{Nonbonding-electron}(\vec{r}') > 0 (Antibonding) \\ \Delta\rho_{Bonding-electron}(\vec{r})\Delta\rho_{Bonding-electron}(\vec{r}') > 0 (Antibonding) \end{cases}$$

(13)

The $V_{bc}(\vec{r}-\vec{r}')$ is deformation charge-bond energy. As such, it induces a transformation

$$\alpha_\sigma^+(\vec{r}) = \sum_{\vec{R}} \psi_{\vec{R}}^*(\vec{r}) a_{\vec{R}\sigma}^+ \equiv \sum_i \psi_{\vec{R}i}^*(\vec{r}) a_{i\sigma}^+ \ .$$

(14)

Then, applied to the coulomb interaction $U_{ii'jj'}$, the transformation (12) and (14) leads to the expansion $\Delta \hat{V}_{ee} = \sum_{ii'jj'} \Delta U_{ii'jj'} a_{i\sigma}^+ a_{i'\sigma'}^+ a_{j\sigma} a_{j'\sigma'}$ where

$$\Delta U_{ii'jj'} = \frac{1}{2} \int d^3r \int d^3r' \psi_{\vec{R}i}^*(\vec{r}) \psi_{\vec{R}j}(\vec{r}) \Delta V(\vec{r}-\vec{r}') \psi_{\vec{R}i'}^*(\vec{r}') \psi_{\vec{R}j'}(\vec{r}').$$

(15)

Where the sum of repeated spin indices is implied, defines the atomic bonding representation of the interaction Hamiltonian.

The combination of the BE model and the BC model is called the BBC model. In the BBC model, atomic bonding and electron states are a unified concept. There are three states of atomic bonding and electron states: antibonding-electron states, nonbonding-electron and bonding-electron states. **Fig. 1** shows a schematic of BBC model.

## 3. Results and discussion

### 3.1 Structural characteristics of defective graphene/BN van der Waals heterostructures

We used QuantumATK tool to generate the initial graphene/BN structure, as shown in **Fig. 2**. The lattice matching of the heterojunction structure is less than 1.14%. The initial structure of the graphene/BN heterojunction had 16 C atoms, eight nitrogen atoms, and eight boron (B) atoms. The lattice parameters are presented in **Table 1**. Note that we labeled each C atom in the graphene layer of the initial graphene/BN heterojunction to facilitate subsequent work. The all defects are manufactured on the graphene layer. They are referred to as Nos. 1 to 16 in **Fig. 2**. We began to manufacture the defects of Nos. 1 to 16 atoms. In the process of constructing the heterojunctions, we obtained a total of 72 different defect structures, which can be seen **in the supporting information**.

We performed DFT calculations for structural optimization. As can be ascertained from **Figs. S1-S4.** and **Table S1**, we calculated the band gap and total energy values for all 72 of the initially manufactured defect structures. However, in the process of optimizing and calculating 72 structures, we found that 30 structures did not successfully converge. Of the remaining 42 structures, there were 13 structures without bandgaps. The five structures we chose to further investigate correspond to those that had the lowest total energies with the two C atoms, three C atoms, four C atoms, five C atoms and six C atoms defect of graphene, **see the supporting information**. The graphene/BN (13,2), graphene/BN (13,2,14), graphene/BN (2,12,14,11), graphene/BN (13,2,14,11,16), and graphene/BN (10,5,6,8,15,13) structures with the lowest energies have defects of C atoms. The five structures evaluated in this study are shown in **Fig. 2**, and the total energies are listed in **Table 2**. We speculate that the defect of atoms caused the structure to become less stable.

Regarding the defect-induced changes in the heterojunction structure (see **Fig. 2**), in the cases of some structures, there was minimal change along the plane inside the heterojunction. This was true for graphene/BN (13,2), graphene/BN (2,12,14,11), and graphene/BN (10,5,6,8,15,13). In the cases of graphene/BN (13,2,14) and graphene/BN (13,2,14,11,16), their planar structures underwent significant changes. Particularly, we found that a flat plane changed to a plane wavy surface of the graphene layer. This change may have altered the stress field of each of the structures. Of course, we can also see that, in the cases of the graphene/BN (2,12,14,11), graphene/BN (13,2,14), and graphene/BN (13,2), after optimization, the original six-membered ring of each of their respective graphene layers was found to have a five-membered ring, or even a three-membered ring.

We also calculated the interlayer interaction energies ($E_I$)[40] of the five heterojunctions, as follows:

$$E_I = E_{total}^{heterostructure} - E_{total}^{BN} - E_{total}^{Graphene}$$

(16)

The calculated results are presented in **Table 3**. A negative interlayer interaction energy indicates that the energetic conditions of the heterostructure were favorable for formation. Specifically, low interlayer interaction energy of the heterostructure is associated with a lower total energy, which yields the formation of a more stable configuration of graphene/BN heterojunctions.

## 3.2 Electronic properties of graphene/boron nitride van der Waals heterostructures with manufactured defects

The results of band gap values for the graphene/BN (13,2), graphene/BN (13,2,14), graphene/BN (2,12,14,11), graphene/BN (13,2,14,11,16), and graphene/BN (10,5,6,8,15,13) heterojunctions are shown in **Fig. 3**. We can see that removing the Nos. 13 and 2 C atoms from the structure resulted in a band gap value of 0.258 eV, when the Nos. 13, 2, and 14 C atoms were removed from the structure, the band gap value was 0.484 eV, when the Nos. 2, 12, 14, and 11 C atoms were defect, the band gap value was 0.507 eV, when the Nos. 13, 2, 14, 11, and 16 C atoms were defect, the band gap value was 0.378 eV. Lastly, the band gap was 0.946 eV when the Nos. 10, 5, 6, 8, 15, and 13 C atoms were defect. Thus, graphene/BN (13,2) heterostructure had the smallest band gap (i.e., 0.258 eV), and graphene/BN (10,5,6,8,15,13) heterostructure had the largest band gap (i.e., 0.946 eV). We infer that the value of the band gap can be controlled by adjusting the number of atomic defect. It is possible to apply this method in the semiconductor manufacturing industry to control the threshold current of materials. It should also be noted that we observed a very rare flat-band phenomenon in the band gap of the graphene/BN (13,2,14,11,16) structure. A distinct horizontal band with negligible dispersion was apparent in the region corresponding to where the Fermi level was close to zero[41].

The defects structure can adjust the electronic dispersion of the semiconductor materials. The distribution of energy is also very important to the physical system, which directly influences the performance of the material and the state of the system. The partial density of states (PDOS) results for the five structures are shown in **Fig. 4**. The electron orbitals of the C atoms were $1s1$, $2s2$, and $2p2$, the orbitals of the B atoms were $1s2$, $2s2$, and $2p3$, the valence electron orbitals of the B atom were $2s2$ and $2p1$. The results in **Fig. 4** show the energy band of the graphene/BN (13,2), graphene/BN (13,2,14), graphene/BN (2,12,14,11), graphene/BN (13,2,14,11,16), and graphene/BN (10,5,6,8,15,13) heterojunctions. During the analysis, we found that $p$ electrons are localized. We also found that five heterojunctions of the $p$ orbital contributed the most energy to the heterojunction, whereas the *s* orbital contribution were very small.

Taking this into consideration with the band gap results, we know that the five heterojunctions are all semiconductors. All five structures were found to have specific energy level shifts, which we know to be influenced by structural defects. As can be seen in **Fig. 5**, the primary peaks of all the structures were in the negative energy region, indicating that the structures were relatively stable. Taking this into consideration with the band gap results, we believe that a higher band gap value corresponds to higher energy of the structure at the Fermi level. Thus, we believe that it may be possible to selectively control the energy level shift and adjust the energy level by manufacturing different defects.

**3.3 Deformation charge density and atomic bonding-electron states**

Here we use the BBC model to analyze electrons and bonding characterization. We believe that the formation of the graphene/BN heterojunction is primarily attributable to the contributions of the charge transfer, atomic bonding and electron states. Note that the electrons in the negative energy region are divergent, whereas those in the positive energy region are convergent. Thus, because deformation charge density maps are used to study electron transfer between atoms, we produced deformation charge density maps to obtain more insight into the charge transfer process for each of the five heterojunctions. The results are shown in **Fig. 6**, where different colors correspond to different electron transfer intensities.

We applied BBC model in charge transfer and electron states, as shown in the deformation charge density graphs, to analyze the bonding-electron states. The results shown in **Fig. 6** describe the electronic states of the atomic bonds in the five defect structures, i.e., the antibonding-electron states and bonding-electron states. These can be discerned from the color scale. The blue area represents an increase in the number of electrons, whereas the red and white areas indicate a decrease in the number of electrons. The color scale also provides information about the electron distribution of the 2D heterojunction. Additionally, an antibonding state indicates a decrease in the electron density and bond energy in this region. Atomic bonding state indicates that the electron density in the region has increased and that the strong bonds have higher energies than the weak bonds.

**Table 4** lists the deformation charge density $\Delta\rho(\vec{r})$ values for the graphene/BN (13,2), graphene/BN (13,2,14), graphene/BN (2,12,14,11), graphene/BN (13,2,14,11,16), and graphene/BN (10,5,6,8,15,13) heterojunctions, as determined via DFT simulation. We use **Eq. 13** and deformation charge density $\Delta\rho(\vec{r})$ that can be calculated the deformation charge-bond energy $V_{bc}(\vec{r}-\vec{r}')$. Note that the deformation charge-bond energy $V_{bc}(\vec{r}-\vec{r}')$ of the crystal lattices with bonding-electron states and antibonding-electron states was determined based on the electron densities of different bonding states. Hence, the type of bond was determined based on the deformation charge density results.

## 4. Conclusions

By manufacturing defects in the graphene layer on the graphene/BN heterojunction, we were able to obtain different defect structures. Additionally, we performed DFT simulations to optimize the different shape of the defective structures, calculate each of their energies, and obtain their band gap. We evaluated the total energy, PDOS, band gap, and deformation charge density for the five most stable structures. We used a combined BBC model to calculate parameter information for the atomic bonding and electronic properties of graphene/BN van der Waals heterojunction with defect characteristics. Moreover, we have found that, by controlling the differences in defect characteristics, the band gap and other band-related structures of the heterojunction can be controlled. If this can be done experimentally, it may become an effective method for the preparation of semiconductor materials of graphene/BN heterojunction. Furthermore, we also observed the existence of a flat-band structure, i.e., a horizontal band structure without dispersion, at the Fermi level. Therefore, defects can cause a horizontal band structure in graphene/BN heterojunction materials.

**Acknowledgements**

The Scientific and Technological Research Program of Chongqing Municipal Education Commission (KJQN201901424), the Chongqing Natural Science Foundation project (cstc2020jcyjmsxmX0524). Dedicated to the 90th birthday of Yangtze Normal

University.

**CRediT authorship contribution statement**

**Jiannan Wang**: performed calculations and performed data analysis. **Liangjing Ge**: wrote the original paper and data curation. **Anlin Deng**: visualization. **Hongrong Qiu**: investigation. **Hanze Li**:investigation.**Yunhu Zhu**: funding acquisition. **Maolin Bo**: revise the paper and developed binding energy and bond-charge (BBC) model. All authors read and contributed to the manuscript.

**Figure and Table Captions:**

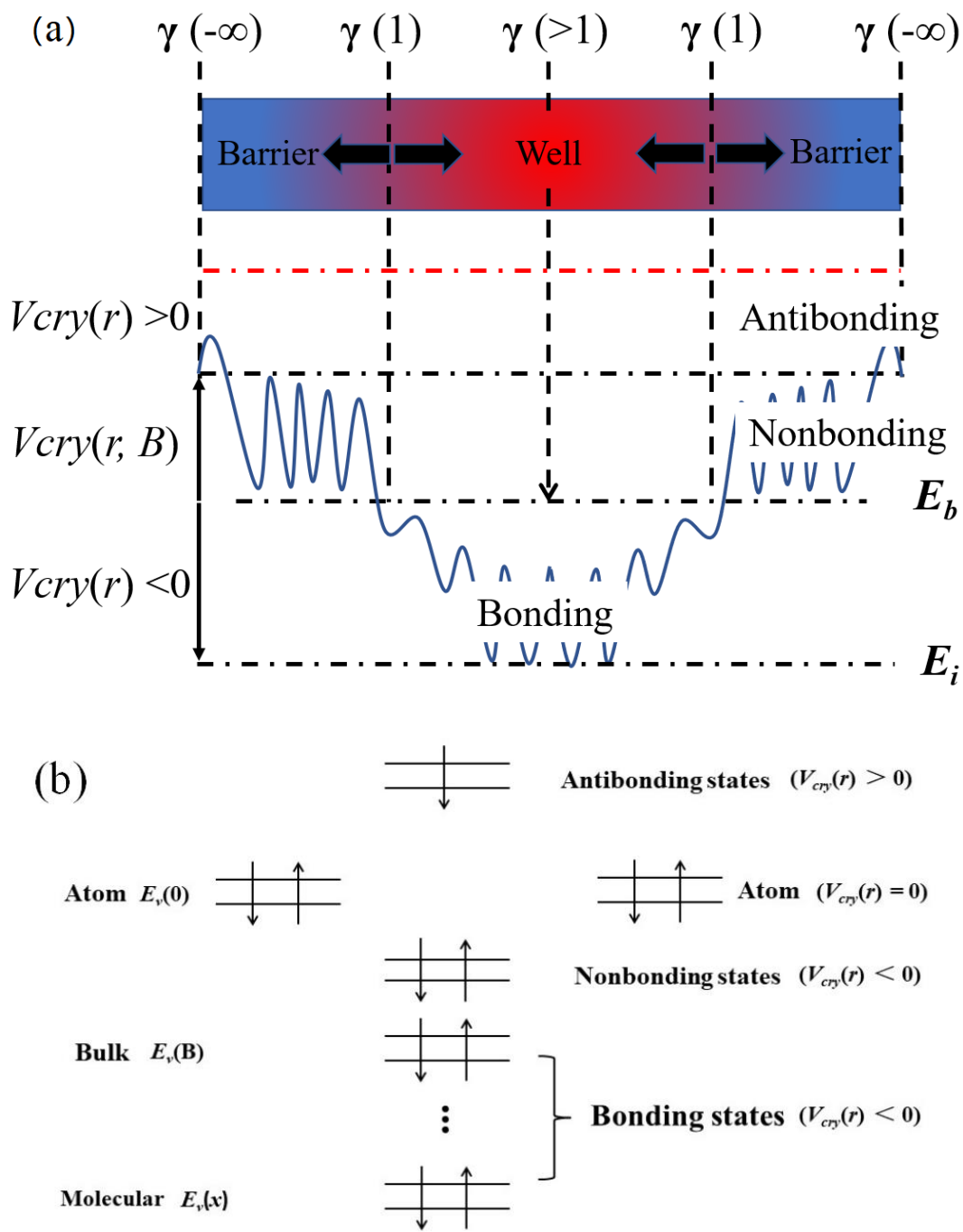

**Figure 1** Schematic showing the (a)BC model and (b)BE model.

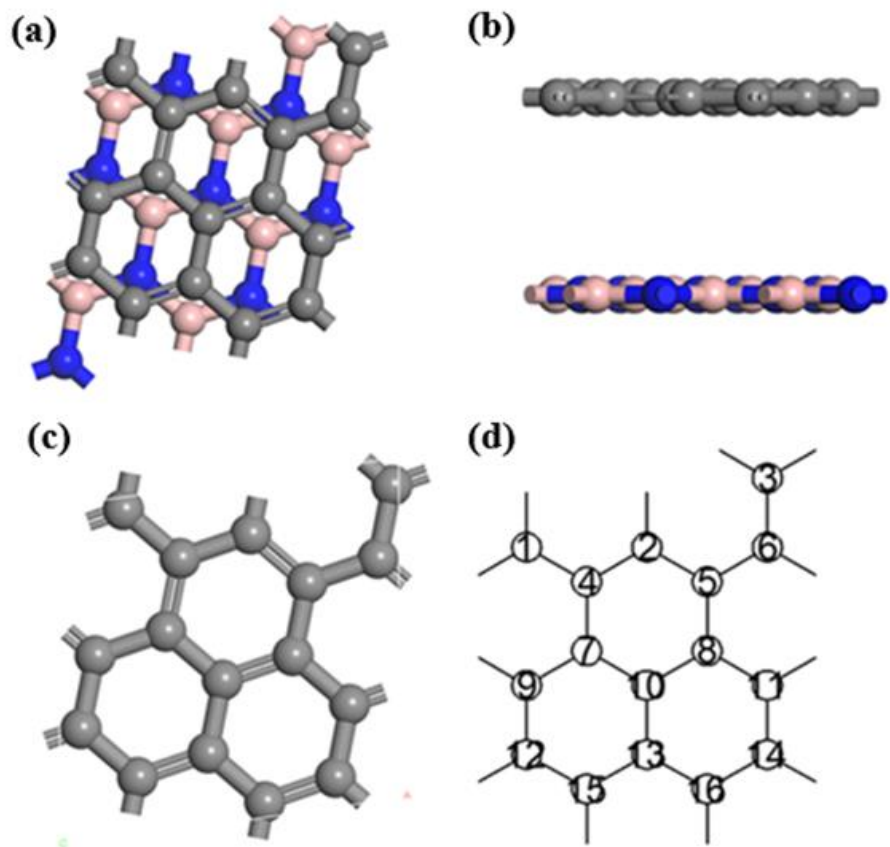

**Figure 2** (a) Top view of the initial Graphene/BN heterojunction structure. (b) Front view of the initial graphene/BN heterojunction structure. (c) Top view of graphene layer in graphene/BN heterojunctions. (d) Corresponding number scheme for the C atoms shown in the graphene layer in (c).

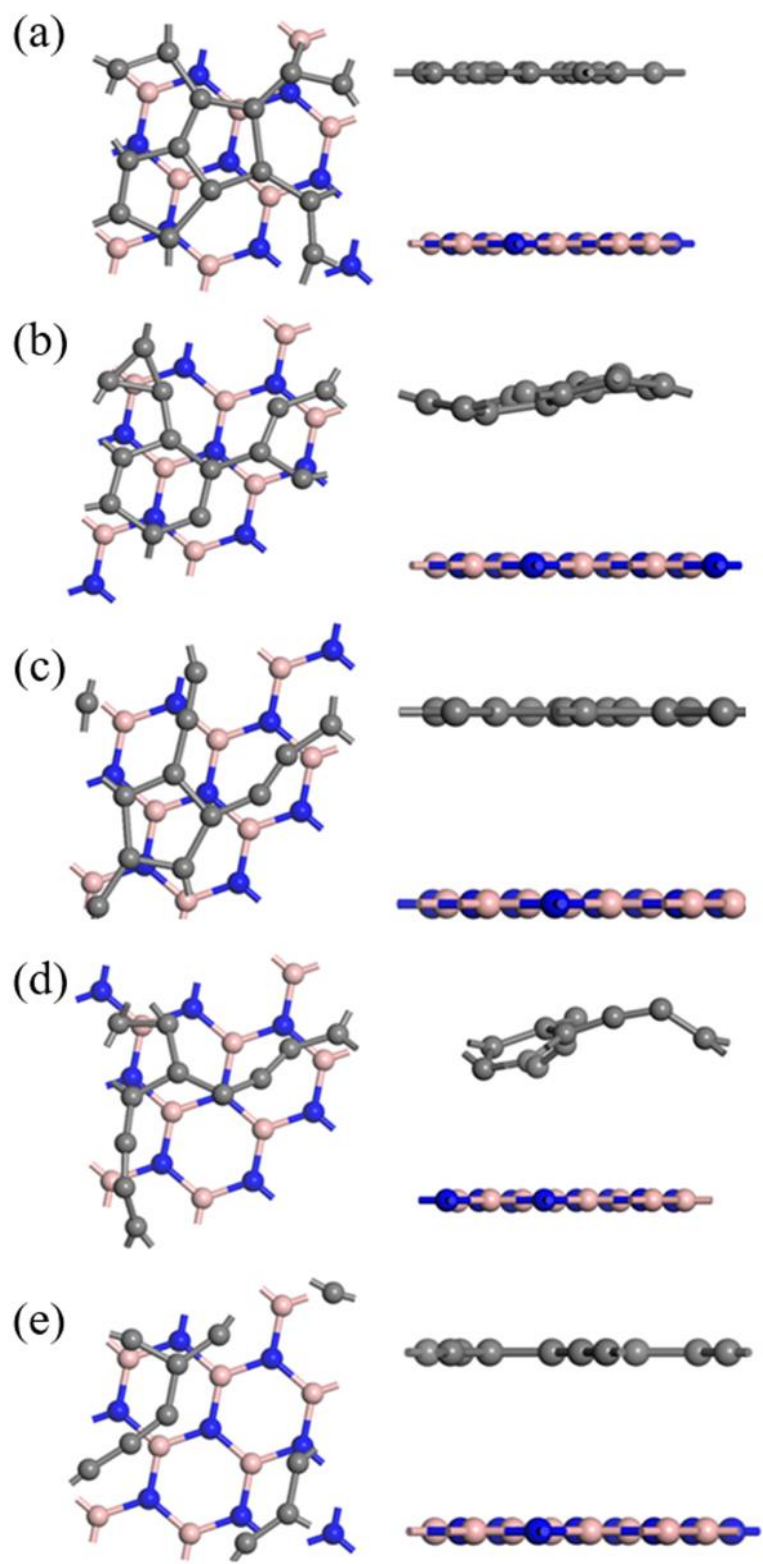

**Figure 3** Five optimized structures selected for analysis. (a)graphene/BN (13,2), (b) graphene/BN (13,2,14), (c)graphene/BN (2,12,14,11), (d)graphene/BN (13,2,14,11,16), (e) graphene/BN (10,5,6,8,15,13) heterojunctions.

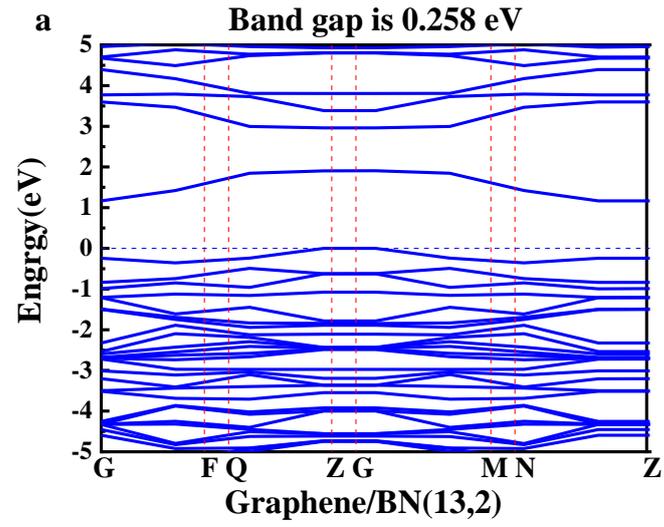

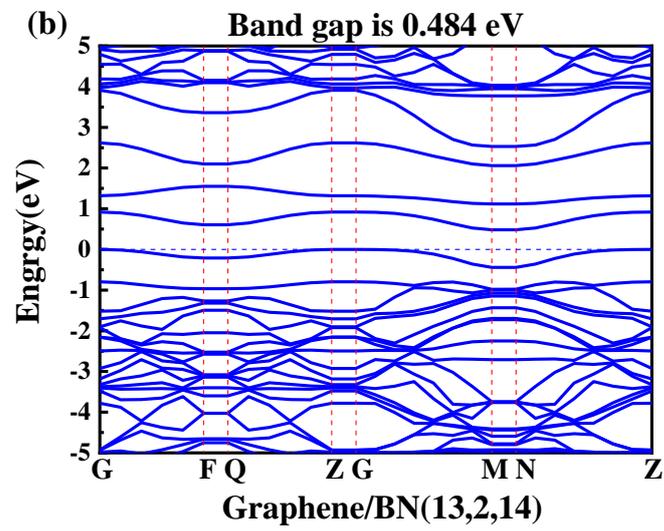

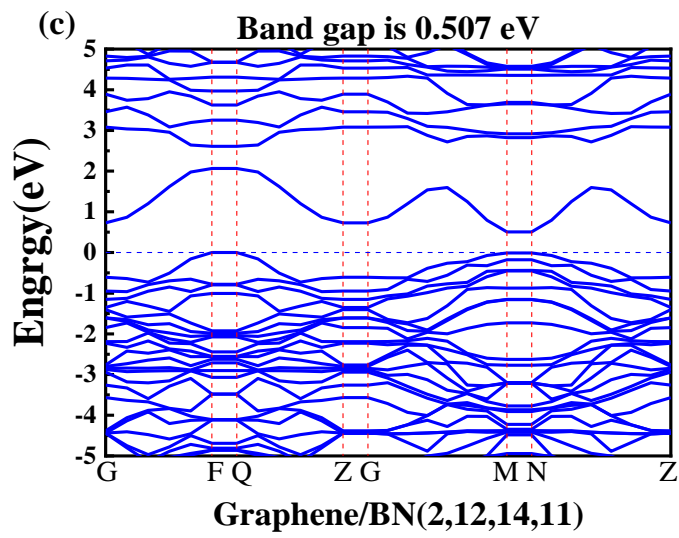

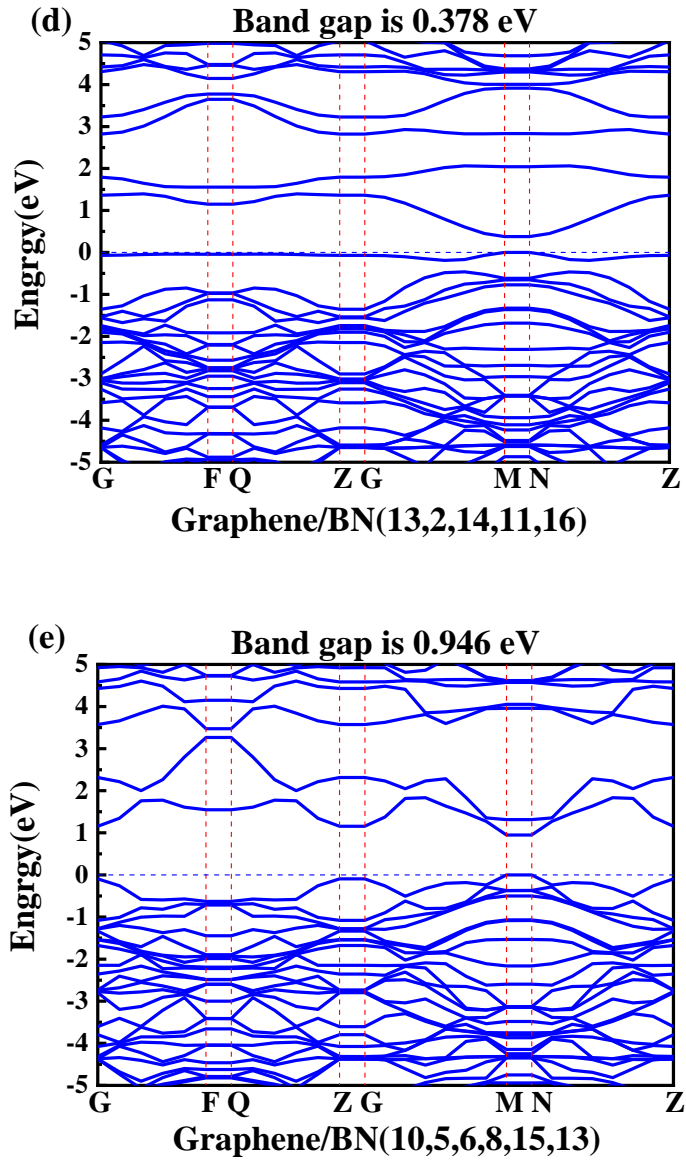

**Figure 4** Band gap results for the (a) graphene/BN (13,2), (b) graphene/BN (13,2,14), (c) graphene/BN (2,12,14,11), (d) graphene/BN (13,2,14,11,16), and (e) graphene/BN (10,5,6,8,15,13) heterojunctions.

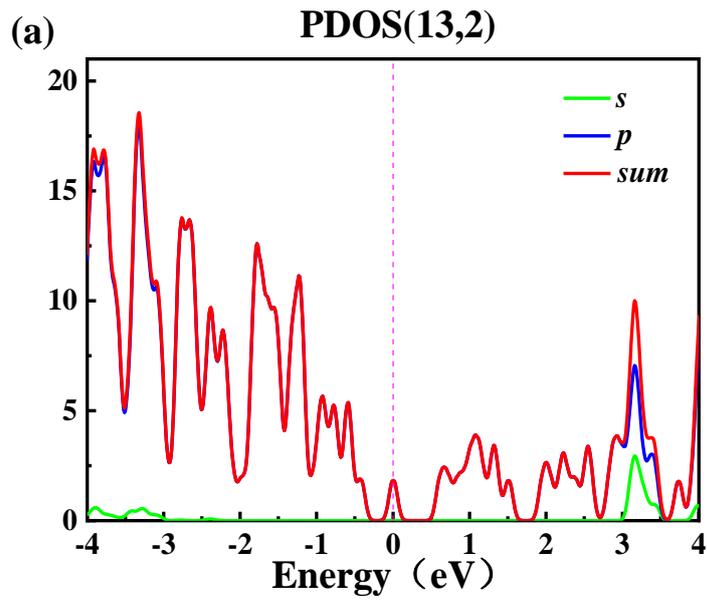

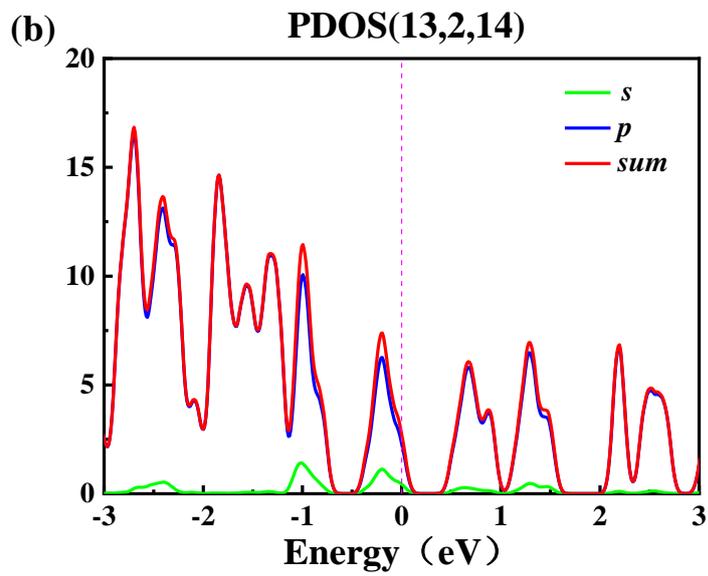

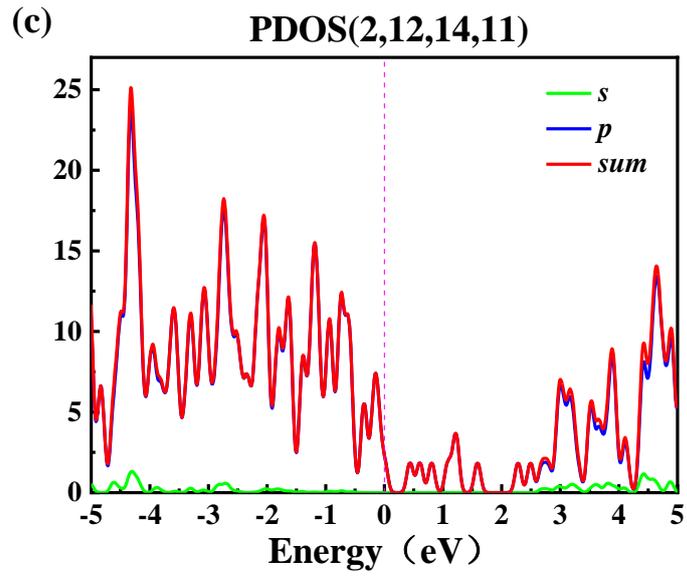

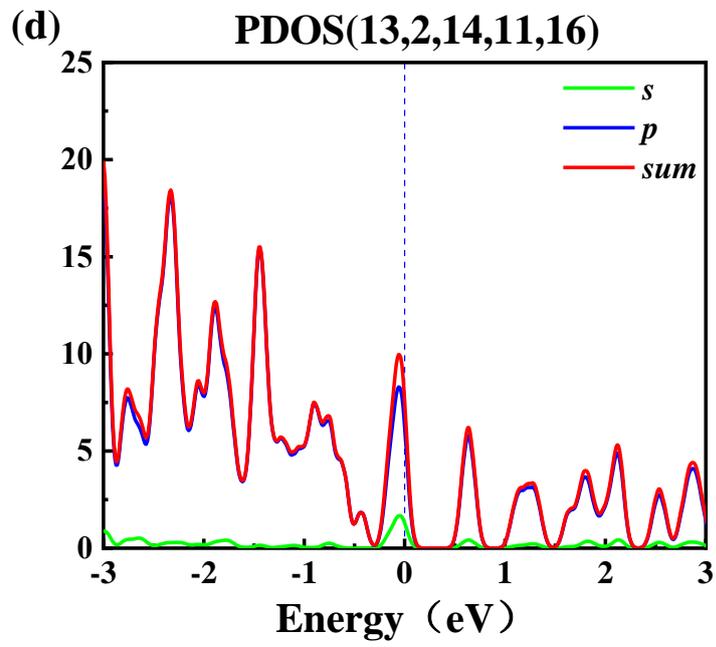

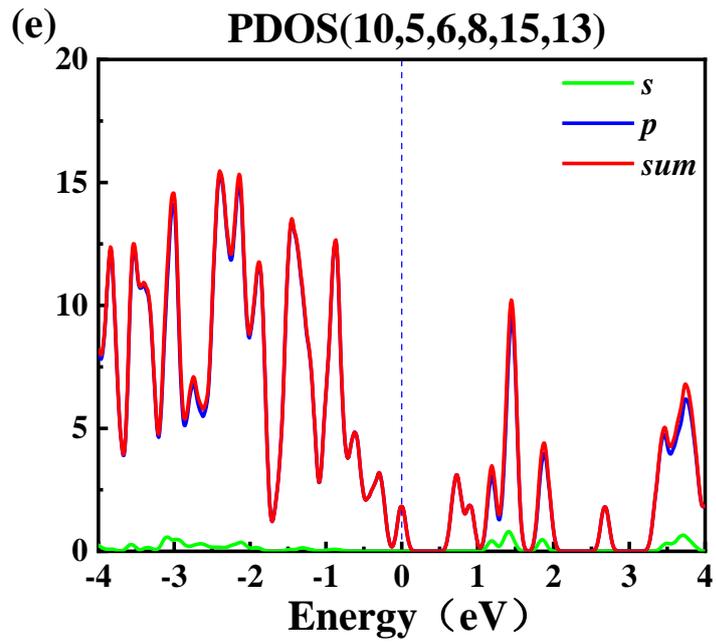

**Figure 5** PDOS results for the (a) graphene/BN (13,2), (b) graphene/BN (13,2,14), (c) graphene/BN (2,12,14,11), (d) graphene/BN (13,2,14,11,16), and (e) graphene/BN (10,5,6,8,15,13) heterojunctions.

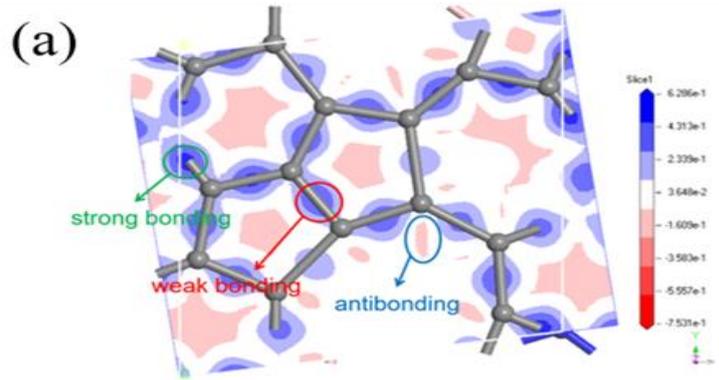
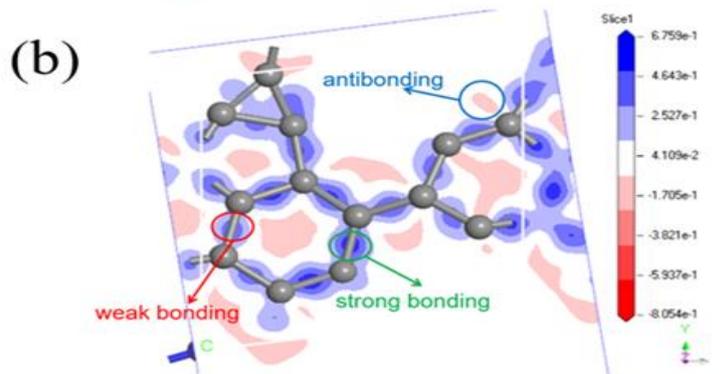
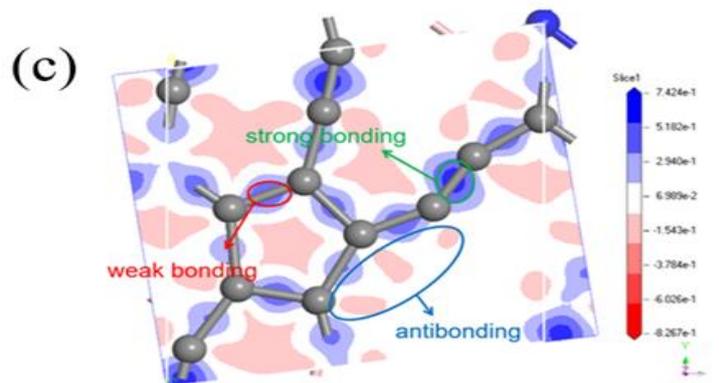
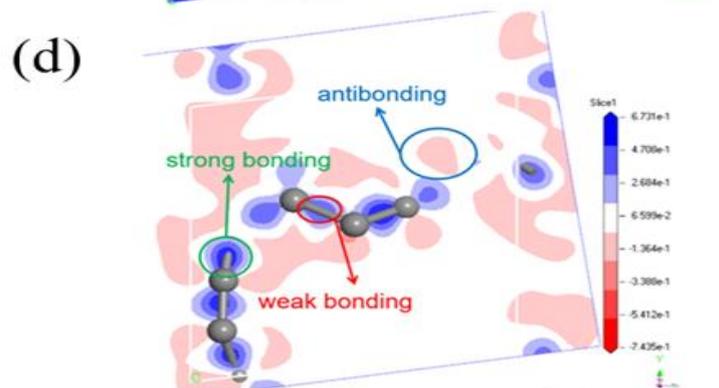
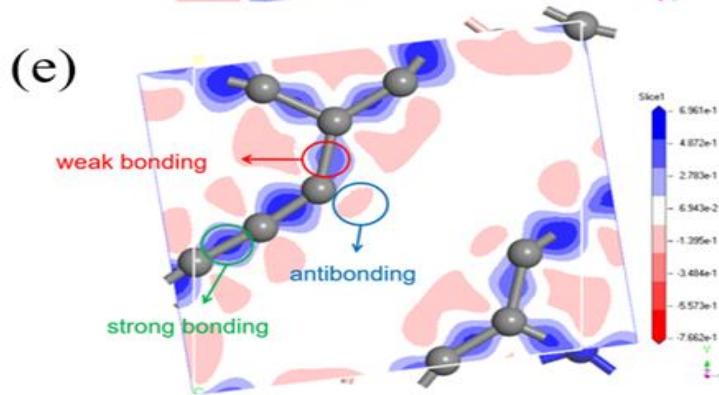

**Figure 6** Deformation charge density maps for the (a) graphene/BN (13,2), (b) graphene/BN (13,2,14), (c) graphene/BN (2,12,14,11), (d) graphene/BN (13,2,14,11,16), and (e) graphene/BN (10,5,6,8,15,13) heterojunctions.

**Table 1** Lattice parameters for the initial graphene/BN heterostructures.

| graphene/BN | Angle | | | Lattice parameters | | | Spacing | k-points |
|---|---|---|---|---|---|---|---|---|
| | α | β | γ | a | b | c | | |
| | 90° | 90° | 82° | 6.645 Å | 6.645 Å | 28.2678Å | 4.187 Å | 4×4×1 |

**Table 2** Total energy and band gap of graphene/BN heterostructures.

| Structure name | Number of defects | Band gap (eV) | Total energy (eV) |
|---|---|---|---|
| graphene/BN (13,2) | 2 | 0.258 | -5157.307 |
| graphene/BN (13,2,14) | 3 | 0.484 | -4995.530 |
| graphene/BN (2,12,14,11) | 4 | 0.507 | -4838.703 |
| graphene/BN (13,2,14,11,16) | 5 | 0.378 | -4679.798 |
| graphene/BN (10,5,6,8,15,13) | 6 | 0.946 | -4524.291 |

**Table 3** Formation energy results for the graphene/BN heterojunctions.

| | $E_{total}^{heterostructure}$ (eV) | | $E_{total}^{BN}$ (eV) | | $E_{total}^{graphene}$ (eV) | $E_I$ (eV) |
|---|---|---|---|---|---|---|
| (13,2) | -5157.30 | | -2957.77 | | -2198.67 | -0.85604 |
| (13,2,14) | -4995.53 | | -2957.77 | | -2036.99 | -0.75804 |
| (2, 12, 14, 11) | -4838.70 | BN | -2957.77 | graphene | -1880.30 | -0.62036 |
| (13, 2, 14, 11, 16) | -4679.79 | | -2957.77 | | -1721.42 | -0.55977 |
| (10, 5, 6, 8, 15, 13) | -4524.29 | | -2957.78 | | -1565.91 | -0.51955 |

**Table 4** The deformation charge density $\Delta\rho(\vec{r})$, and deformation charge-bond energy $V_{bc}(\vec{r}-\vec{r}')$, results for the van der Waals heterojunctions, as obtained by calculating the bond charge model.

($\varepsilon_0 = 8.85\times10^{-12}C^2N^{-1}m^{-2}, e=1.60\times10^{-19}C, |\vec{r}-\vec{r}'|=d_{ij}/2=1.49\,\text{Å}$)

| graphene/BN | (13,2) | (13,2,14) | (2,12,14,11) | (13,2,14,11,16) | (10,5,6,8,15,13) |
|---|---|---|---|---|---|
| $\Delta\rho^{hole-electron}(\vec{r})\,(e/\text{Å}^3)$ | -0.7531 | -0.8054 | -0.8267 | -0.7435 | -0.7662 |
| $\Delta\rho^{Nonbonding-electron}(\vec{r}')\,(e/\text{Å}^3)$ | 0.4313 | 0.4643 | 0.5182 | 0.4708 | 0.4872 |
| $\Delta\rho^{Bonding-electron}(\vec{r}')\,(e/\text{Å}^3)$ | 0.6286 | 0.6759 | 0.7424 | 0.6731 | 0.6961 |
| $\Delta\rho^{Antibonding-electron}(\vec{r}')\,(e/\text{Å}^3)$ | 0.1705 | 0.1705 | 0.4213 | 0.3350 | 0.3530 |
| $V_{bc}^{Weak\,bonding}(eV)$ | -0.5365 | -0.6177 | -0.7076 | -0.5782 | -0.5365 |
| $V_{bc}^{Strong\,bonding}(eV)$ | -0.7819 | -0.89915 | -1.0137 | -0.8266 | -0.7819 |
| $V_{bc}^{Antibonding}(eV)$ | 0.2001 | 0.2268 | 0.2107 | 0.1675 | 0.2001 |

# Supporting Information for Publication

# Atomic bonding and electrical characteristics of two-dimensional graphene/boron nitride van der Waals heterostuctures with manufactured defects

Jiannan Wang, Liangjing Ge, Anlin Deng, Hongrong Qiu, Hanze Li, Yunhu Zhu,

Maolin Bo*

Key Laboratory of Extraordinary Bond Engineering and Advanced Materials Technology

(EBEAM) of Chongqing, Yangtze Normal University, Chongqing 408100, China

*Corresponding Author: E-mail: bmlwd@yznu.edu.cn (Maolin Bo)

**Optimal geometric structures of defective graphene/BN van der Waals heterostuctures**

When the graphene layer loses 1 C atom (the carbon atom number is 10), as shown in the **Figure S1**. The 10th C atom is absent in this structure, so we named the structure Graphene/BN (10). For another example in the Figure S2, the 2nd C atom and the 13th C atom are absent, so we named this structure Graphene/BN (13 2) heterojunction, the complete structure of is shown in the Figure S3. As shown in the Figure S4, this is a schematic diagram of all the 72 types of graphene layers without repeating lattice. However, in the process of optimizing and calculating 72 structures, we found that 30 structures did not successfully converge. Of the remaining 42 structures, there were 13 structures without bandgaps.(Graphene/BN(1), Graphene/BN(15,7,11),Graphene/BN(2,5,10,8),Graphene/BN(2,12,14,9),Graphene/BN(10,5,6,8),Graphene/BN(14,10,1,6),Graphene/BN(10.5.6.8.15),Graphene/BN(14,10,1,6,3),Graphene/BN(15,7,11,12,6),Graphene/BN(13,2,16,12,1),Graphene/BN(3,14,9,2,16), Graphene/BN(2,5,10,8,4,7,11), Graphene/BN (2,5,10,8,4,7)) Finally, we get the remaining 29 optimized structures with band gaps, as shown in the Table S1.

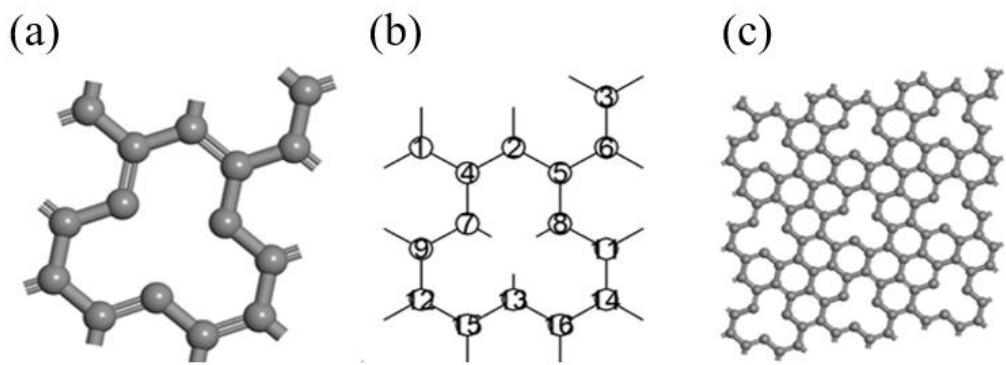

**Figure S1** (a) Graphene layer structure of Graphene/BN (10) (b) Graphene/BN (10) C atom number (C) Graphene/BN (10) Periodic structure of graphene layers.

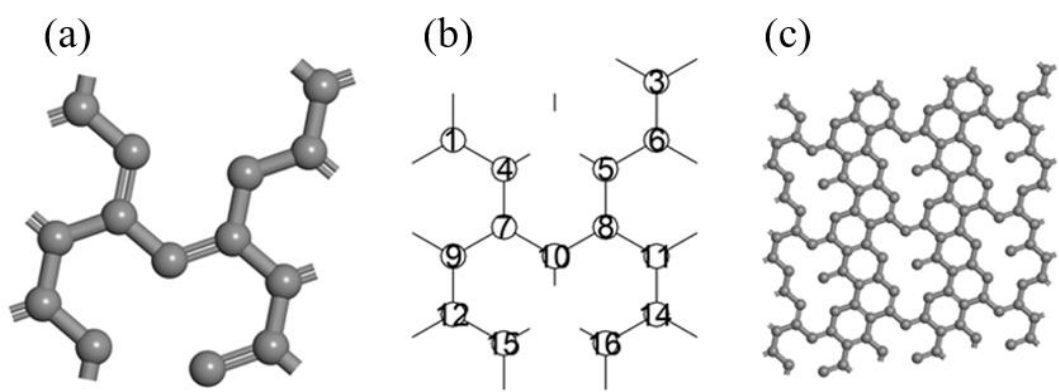

**Figure S2** (a) Graphene layer structure of Graphene/BN (13, 2) (b) Graphene/BN (13, 2) C atom number (c) Graphene/BN (13, 2) Periodic structure of graphene layer.

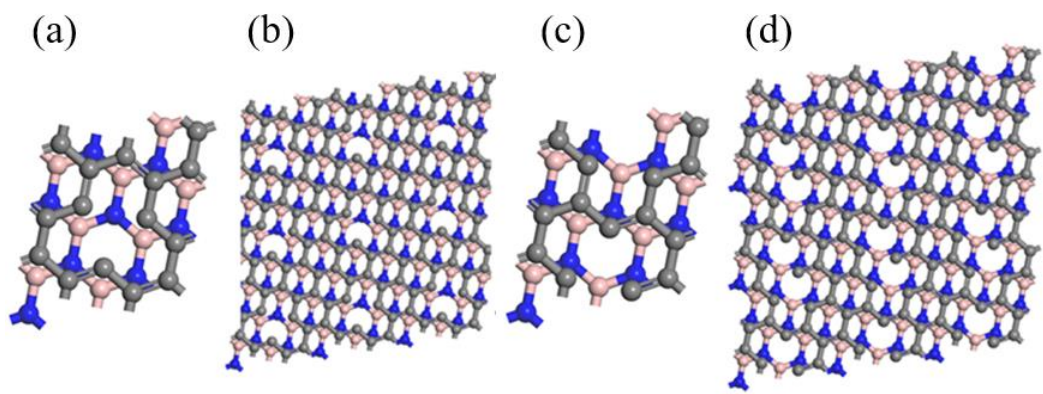

**Figure S3** (a) Graphene/BN (10) structure (b) Graphene/BN (10) d periodic structure (c) Graphene/BN (13, 2) structure (d) Graphene/BN (13, 2) Periodic structure.

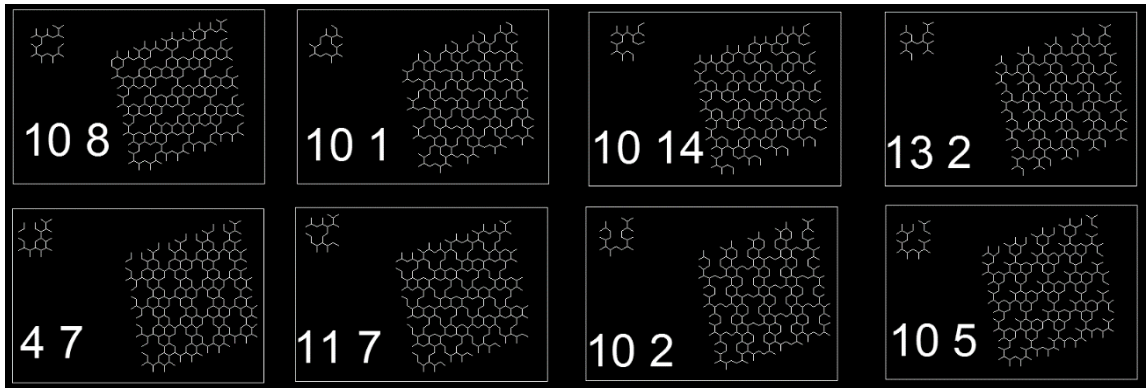
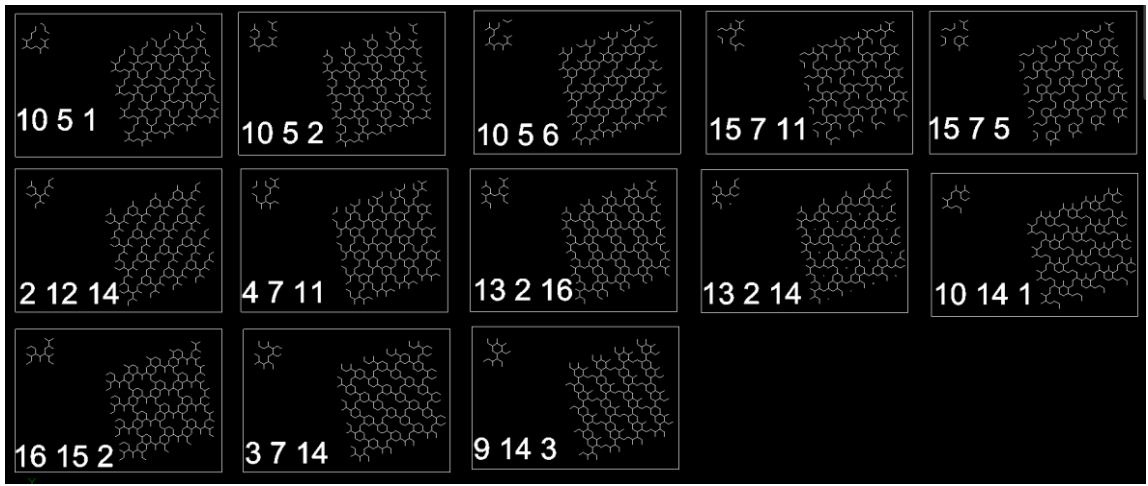
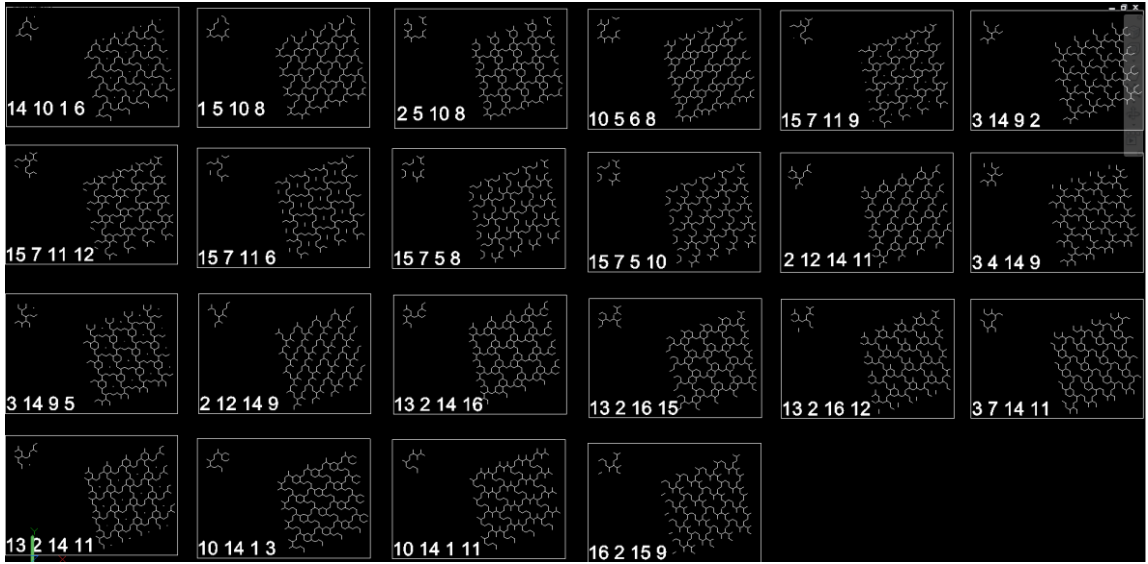

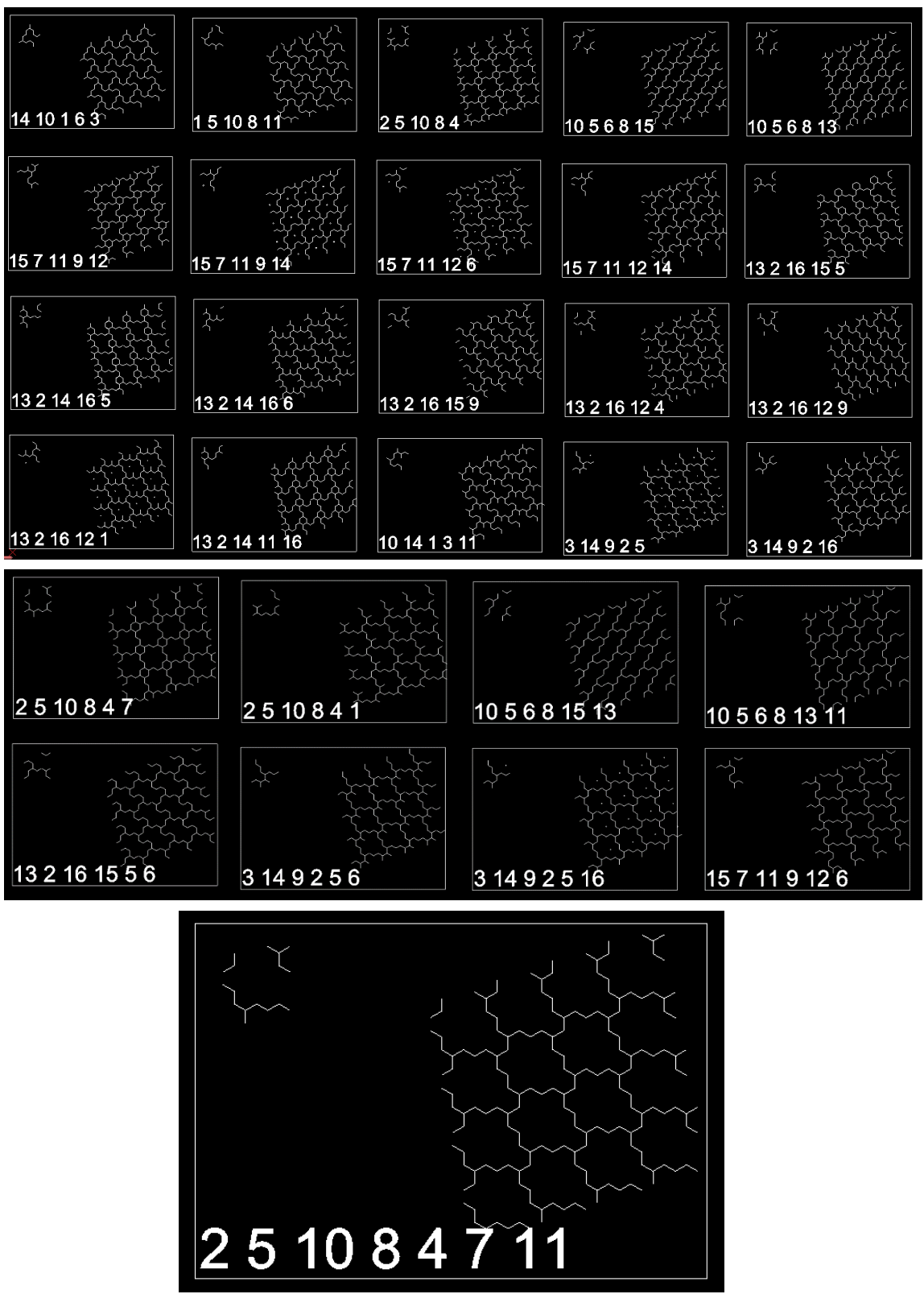

**Figure S4** Schematic diagram of 72 graphene layers with non-repetitive structures.

Table S1 Band gap and energy values of the remaining 29 structures.

| Number of atoms absent | Number of electronic absent | Total energy(eV) | Band gap(eV) |
|---|---|---|---|
| 2 | (10,2) | -5154.666 | 0.167 |
| | (13,2) | -5157.307 | 0.258 |
| | (8,1) | -5157.39 | 0.343 |
| 3 | (15,7,5) | -4994.521 | 0.319 |
| | (3,7,14) | -4994.065 | 0.207 |
| | (13,2,14) | -4995.53 | 0.484 |
| 4 | (1,5,10,8) | -4837.809 | 0.921 |
| | (2,12,14,11) | -4838.703 | 0.507 |
| | (10,14,1,3) | -4835.61 | 0.332 |
| | (13,2,14,16) | -4837.292 | 0.971 |
| | (13,2,16,12) | -4837.904 | 1.273 |
| | (15,7,11,12) | -4835.116 | 0.631 |
| 5 | (2.5.10.8.4) | -4680.25 | 0.353 |
| | (10,5,6,8,13) | -4680.314 | 0.627 |
| | (13,2,16,15,9) | -4679.513 | 1.178 |
| | (13,2,16,12,4) | -4679.754 | 0.147 |
| | (15,7,11,12,14) | -4678.552 | 1.037 |
| | (13,2,16,12,9) | -4679.438 | 1.177 |
| | (13,2,14,11,16) | -4680.286 | 0.38 |
| | (3,14,9,2,5) | -4678.397 | 0.056 |
| | (10,14,1,3,11) | -4679.532 | 0.973 |
| 6 | (2,5,10,8,4,1) | -4522.251 | 1.644 |
| | (10,5,6,8,15,13) | -4524.295 | 1.245 |
| | (3,14,9,2,5,6) | -4523.319 | 1.168 |
| | (10,5,6,8,13,11) | -4524.036 | 1.169 |
| | (13,2,16,15,5,6) | -4523.504 | 0.714 |

| | | |
|---|---|---|
| (3,14,9,2,5,6) | -4523.319 | 1.168 |
| (3,14,9,2,5,16) | -4524.163 | 2.15 |
| (10,5,7,9,12,6) | -4522.022 | 0.66 |